\documentclass[journal]{IEEEtran}
\usepackage{amsmath,amsfonts,amssymb,amsbsy,bm,paralist,theorem,color}
\usepackage{graphicx,relsize}
\usepackage{algorithm}
\usepackage{algpseudocode}
\usepackage{multicol}
\usepackage{multirow}
\usepackage{subcaption}
\usepackage{cite}
\usepackage{hyperref}
\usepackage{cuted}
\usepackage{pifont}
\usepackage{comment}
\usepackage{xcolor}
\usepackage{flushend}

\graphicspath{{fig/}}

\definecolor{orange}{RGB}{255,107,0}
\definecolor{green}{RGB}{0,160,20}

\begin{document}
\title{Fairness-Oriented Optimization of NOMA-Enabled Pinching-Antenna Systems Under Blockage and Imperfect CSI}

\author{Zhehang Ye, Ximing Xie,~\IEEEmembership{Member,~IEEE}, Hao Qin,~\IEEEmembership{Member,~IEEE}, Xingqi Zhang,~\IEEEmembership{Senior Member,~IEEE}, and Yuanwei~Liu,~\IEEEmembership{Fellow,~IEEE}
\thanks{Zhehang Ye and Hao Qin are with the School of Electronics and Information Engineering, Sichuan University, Chengdu, China (hao.qin@scu.edu.cn).}
\thanks{Ximing Xie is with the Department of Electrical and Computer Engineering, Western University, London, ON N6A 3K7, Canada (xxie269@uwo.ca).}
\thanks{Xingqi Zhang is with the Department of Electrical and Computer Engineering, University of Alberta, Canada T6G 2H5 (e-mail: xingqi.zhang@ualberta.ca).}
\thanks{Yuanwei Liu is with the Department of Electrical and Electronic Engineering, The University of Hong Kong, Hong Kong (e-mail: yuanwei@hku.hk).}
}

\maketitle

\begin{abstract}
The pinching-antenna system (PASS) has been proposed as a promising solution for mitigating line-of-sight (LoS) blockages by dynamically repositioning pinching antennas (PAs) along a dielectric waveguide. This paper develops a fairness-oriented downlink design for a non-orthogonal multiple access (NOMA)-enabled PASS, where the longitudinal placement of PAs and the NOMA power allocation coefficients are jointly optimized to maximize the minimum user signal-to-interference-plus-noise ratio (SINR) across all users under transmit power and waveguide constraints. A soft-blockage channel model incorporating waveguide attenuation and imperfect channel state information (CSI) is developed. To ensure the feasibility of successive interference cancellation under CSI uncertainty, a conservative SINR evaluation framework is proposed. The resulting non-convex max–min SINR optimization problem is efficiently solved using a tailored particle swarm optimization (PSO) algorithm. Numerical results demonstrate that the proposed design improves the minimum user SINR by approximately 7-10 dB compared with fixed-antenna systems and non-robust optimization baselines under moderate blockage and imperfect CSI.
\end{abstract}


\IEEEpeerreviewmaketitle


\section{Introduction}
Recently, the pinching-antenna system (PASS) has been proposed as an effective approach to mitigate large-scale fading and line-of-sight (LoS) blockages \cite{suzuki2022pinching}. A typical PASS consists of dielectric waveguides equipped with low-cost pinching antennas (PAs), which enable controlled radiation of electromagnetic waves. By dynamically adjusting the positions of PAs along the waveguide, PASS can reconfigure radiation points to bypass obstacles and establish LoS links, making it particularly suitable for blockage-rich propagation environments \cite{11036558}. 

Since the pinching antennas on a single waveguide share a common transmit signal, PASS is inherently well suited to non-orthogonal multiple access (NOMA) for the simultaneous service of multiple users \cite{10945421}. Existing studies on NOMA-enabled PASS have demonstrated its performance advantages over conventional fixed-antenna systems. For example, \cite{10912473} examined pinching antenna (PA) activation and selection for downlink NOMA to better exploit the waveguide-mediated spatial degrees of freedom. Practical deployment for downlink NOMA was then addressed via a low-complexity PA placement design \cite{11016750}. Building on these, joint optimization of PA positions and NOMA power allocation coefficients was formulated for sum rate improvement and quality of service (QoS) provisioning \cite{11029492,xu2025qosawarenomadesigndownlink}. Transmit power minimization was also investigated as an optimization objective, while ensuring the feasibility of successive interference cancellation (SIC) \cite{11186151}. More recent studies have extended NOMA-enabled PASS beyond power and placement design to include joint beamforming and system-level resource allocation \cite{gan2025jointbeamformingnomaassisted,xue2025resourceallocationpinchingantennasystems}. 

However, existing studies typically assume perfect channel state information (CSI) and do not address fairness-oriented optimization. To fill this gap, this paper maximizes the minimum user signal-to-interference-plus-noise ratio (SINR) under imperfect CSI, thereby ensuring fairness among multiple downlink NOMA users. We first develop a channel model that incorporates soft blockage and imperfect CSI. Accordingly, we introduce a conservative SINR evaluation to guarantee SIC feasibility under CSI errors. With this conservative SINR in place, we formulate a max–min SINR optimization problem that jointly optimizes PA positions and NOMA power allocation coefficients. To solve it, we propose a tailored particle swarm optimization (PSO) algorithm for the resulting non-convex design. Numerical results demonstrate that the proposed design significantly improves the minimum user SINR compared with fixed-antenna systems and non-robust optimization baselines under moderate blockage and imperfect CSI.


\section{System Model and Problem Formulation}
In this section, we first introduce the downlink NOMA-enabled PASS and the associated channel model. We then formulate a max–min SINR optimization problem to jointly optimize the PA positions and the NOMA power allocation coefficients.

\subsection{System Model}
\label{system_description}
\begin{figure}[t]
  \centering
  \includegraphics[width=0.9\linewidth]{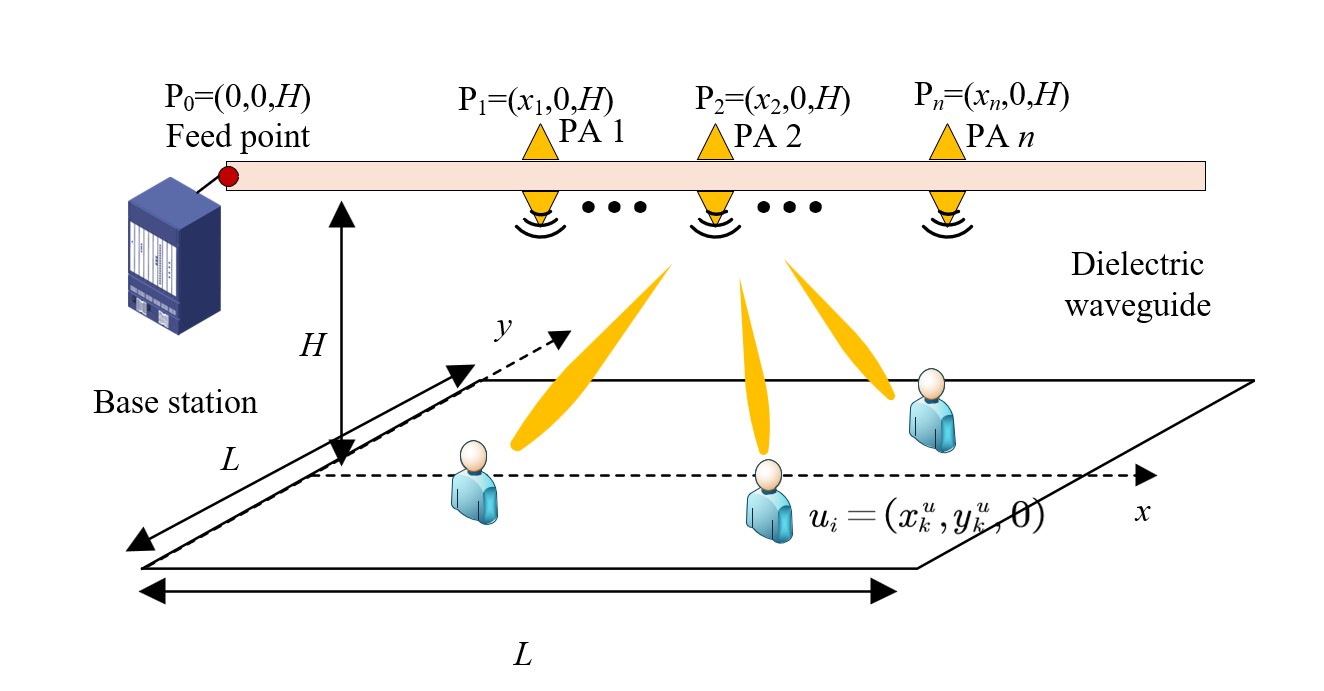}
  \caption{The downlink NOMA-enabled PASS.}
  \label{fig:pass}
\end{figure}

As shown in Fig.~\ref{fig:pass}, the downlink NOMA-enabled PASS consists of one base station (BS) equipped with a dielectric waveguide of length $L$ to serve $K$ single-antenna users. We assume that $N$ PAs are deployed on the waveguide. Each PA taps the guided wave and radiates it into free space. Let the waveguide axis be the $x$-axis with span $[0,L]$.
The $n$-th PA is located at
\begin{equation}
\mathbf{p}_n = [x_n,\, 0 ,\, H]^T,\quad n=1,\ldots,N,
\end{equation}
and we collect the $x$-coordinates of all the PAs as follows:
\begin{equation}
\mathbf{x}_{\!p} = [x_1,\ldots,x_N]^T \in \mathbb{R}^N .
\end{equation}
The PAs are installed at a fixed height $H$. Since all the PAs must lie on the waveguide, they should satisfy a minimum spacing:
\begin{align}
0 \le x_n \le L,\quad & n=1,\ldots,N,
\label{eq:geom_box_pn}\\
|x_n-x_{n-1}| \ge d_{\min},\quad & n=2,\ldots,N,
\label{eq:min_spacing_pn}
\end{align}
where $d_{\min}$ is the minimum allowable separation between adjacent PAs due to physical size and installation constraints.

User $k$ is located at
\begin{equation}
\mathbf{u}_k = [x_k^{(u)},\, y_k^{(u)},\, 0]^T,\quad k=1,\ldots,K,
\end{equation}
where $z=0$ represents the ground plane. We assume that $y_k^{(u)} > 0 $ so that users are in the half-space in front of the waveguide. Although different users may have slightly different effective heights, this variation is small compared with the horizontal distance variations in typical indoor deployments. Therefore, we model the vertical separation as approximately constant and focus on the horizontal coordinates, which constitute the dominant source of geometric variation.

\subsection{Channel and Signal Model}
According to \cite{10945421}, LoS links are much stronger than NLoS links. As a result, we only consider the LoS component in the channel model. We first model the power attenuation along the waveguide as a function of the guided-wave distance from the feed to PA $n$. A common choice is an exponential loss model,
\begin{equation}
\rho_{\mathrm{wg}}(x_n)=10^{-\kappa x_n/10},
\end{equation}
where $\kappa$ is the waveguide loss coefficient in dB/m. Therefore, the deterministic LoS component is modeled as
\begin{equation}
\begin{aligned}
g^{\mathrm{LoS}}_{k,n}
&=
\sqrt{\rho_{\mathrm{wg}}(x_n)}\,
\frac{\lambda}{4\pi r_{k,n}}\,
e^{\left(-j\frac{2\pi}{\lambda}r_{k,n}\right)},
\end{aligned}
\label{eq:channel_los}
\end{equation}
where $r_{k,n} = \|\mathbf{u}_k-\mathbf{p}_n\|_2$ denotes the distance between user $k$ and PA $n$. Meanwhile, we use a soft-blockage factor $b_{k,n}\in[\beta,1]$:
\begin{equation}
b_{k,n}
=
\beta+(1-\beta)\!\left(1-e^{-\alpha d_{k,n}}\right),
\label{eq:soft_blockage}
\end{equation}
where $d_{k,n}$ is the minimum distance from the PA-user line segment to the nearest obstacle surface,
$\alpha>0$ is the attenuation rate, and $\beta\in(0,1]$ denotes the minimum transmission coefficient
(the residual fraction of the LoS component that remains under severe blockage).
Smaller $d_{k,n}$ implies stronger obstruction and thus lower transmission (closer to $\beta$), while larger $d_{k,n}$ yields $b_{k,n}\to 1$.
Then, the channel between user $k$ and PA $n$ is given by
\begin{equation}
\begin{aligned}
g_{k,n}
&=
b_{k,n} g^{\mathrm{LoS}}_{k,n}.
\end{aligned}
\label{eq:channel}
\end{equation}
\par

Let $s_k$ denote the signal for user $k$, satisfying $\mathbb{E}\{|s_k|^2\}=1$.
The BS transmits a superposition signal with total transmit power budget $P_T$:
\begin{equation}
s_{\mathrm{tx}} = \sum_{k=1}^{K}\sqrt{p_k}\,s_k,
\qquad p_k=\alpha_k P_T,
\end{equation}
where $p_k$ denotes the transmit power allocated to user $k$ and
$\boldsymbol{\alpha}=[\alpha_1,\ldots,\alpha_K]^T$ satisfies the simplex constraint
\begin{equation}
\alpha_k \ge 0,
\qquad
\sum_{k=1}^{K}\alpha_k \le 1 .
\label{eq:noma_alpha}
\end{equation}
All PAs are driven by a common feed and radiate the same baseband waveform $s_{\mathrm{tx}}$ with different phase shift,
\begin{equation}
s_{n} = s_{\mathrm{tx}} e^{-j\frac{2 \pi}{\lambda_g} x_n},\quad n=1,\ldots,N ,
\label{eq:pa_excitation_equal}
\end{equation} 
where $\lambda_g$ denotes the waveguide wavelength. Therefore, the received signal of user $k$ is
\begin{align}
    y_k &= \sum\limits_{n=1}^N  g_{k,n} s_n + n_k \nonumber \\
    & = \sum\limits_{n=1}^N \left(b_{k,n}\sqrt{\rho_{\mathrm{wg}}(x_n)}\, \frac{\lambda}{4\pi r_{k,n}}\, e^{\left(-j\frac{2\pi}{\lambda}r_{k,n} - j\frac{2 \pi}{\lambda_g} x_n\right)}\right) s_{\mathrm{tx}} + n_k,
\end{align}
where $n_k\sim\mathcal{CN}(0,\sigma^2)$ denotes the additive white Gaussian noise (AWGN). The effective channel of user $k$ is 
\begin{equation}
    h_k(\mathbf{x}_{\!p}) = \sum\limits_{n=1}^N \left(b_{k,n}\sqrt{\rho_{\mathrm{wg}}(x_n)}\, \frac{\lambda}{4\pi r_{k,n}}\, e^{\left(-j\frac{2\pi}{\lambda}r_{k,n} - j\frac{2 \pi}{\lambda_g} x_n\right)}\right). \label{effective_channel}
\end{equation}
To account for the imperfect CSI, we adopt a relative bounded CSI error model
\begin{equation}
\hat h_k = h_k + e_k, \qquad |e_k| \le \varepsilon |h_k|, \qquad \varepsilon\in[0,1), \label{eq:csi_error}
\end{equation}
where $e_k$ and $\varepsilon$ denote the channel estimation error and the relative error bound, respectively. \par


\subsection{SIC Feasibility under Imperfect CSI}
To analyze the SIC feasibility under imperfect CSI, we first recall the standard SIC feasibility condition under perfect CSI, and then explain why a conservative ordering is needed under \eqref{eq:csi_error}.
We assume that channels are ordered as
\begin{equation}
|h_1(\mathbf{x}_{\!p})|^2 \le \cdots \le |h_K(\mathbf{x}_{\!p})|^2. \label{eq:ordering_true}
\end{equation}
User $k$ decodes $s_k$ after canceling $\{s_1,\ldots,s_{k-1}\}$, so its SINR is
\begin{equation}
\mathrm{SINR}^{\mathrm{true}}_k(\mathbf{x}_{\!p},\boldsymbol{\alpha})
=
\frac{\alpha_k P_T |h_k(\mathbf{x}_{\!p})|^2}{
\sum_{j=k+1}^{K}\alpha_j P_T |h_k(\mathbf{x}_{\!p})|^2 + \sigma^2 } .
\label{eq:sinr_true_k}
\end{equation}
When user $j>k$ decodes $s_k$ for SIC, its SINR is
\begin{equation}
\mathrm{SINR}^{\mathrm{true}}_{k\rightarrow j}(\mathbf{x}_{\!p},\boldsymbol{\alpha})
=
\frac{\alpha_k P_T |h_j(\mathbf{x}_{\!p})|^2}{
\sum_{i=k+1}^{K}\alpha_i P_T |h_j(\mathbf{x}_{\!p})|^2 + \sigma^2 } .
\label{eq:sinr_true_k_to_j}
\end{equation}
The SIC requirement is $\mathrm{SINR}^{\mathrm{true}}_{k\rightarrow j}\ge \mathrm{SINR}^{\mathrm{true}}_k$ for all $j>k$.
For $\alpha_k>0$, cancel $\alpha_k P_T$ and define $S_k\triangleq\sum_{i=k+1}^{K}\alpha_i\ge 0$ and $\eta\triangleq\sigma^2/P_T>0$:
\begin{align}
\frac{|h_j|^2}{|h_j|^2 S_k + \eta}
&\ge
\frac{|h_k|^2}{|h_k|^2 S_k + \eta},
\label{eq:sic_reduce_a}\\
|h_j|^2\big(|h_k|^2 S_k + \eta\big)
&\ge
|h_k|^2\big(|h_j|^2 S_k + \eta\big),
\label{eq:sic_reduce_b}\\
|h_j|^2 &\ge |h_k|^2,\quad \forall j>k.
\label{eq:sic_reduced}
\end{align}
Thus, \eqref{eq:ordering_true} guarantees SIC feasibility under the perfect channels.\par

In practice, the decoding order must be constructed from the estimates $\{\hat h_k\}$, which may not preserve \eqref{eq:ordering_true} under \eqref{eq:csi_error}.
From \eqref{eq:csi_error} and the triangle inequality, for $\varepsilon<1$,
\begin{equation}
\frac{|\hat h_k|}{1+\varepsilon}
\le
|h_k|
\le
\frac{|\hat h_k|}{1-\varepsilon}.
\label{eq:h_bounds}
\end{equation}
Therefore, a sufficient condition to guarantee $|h_j|\ge |h_k|$ for all channels in the uncertainty set is
\begin{equation}
\frac{|\hat h_j|}{1+\varepsilon}
\ge
\frac{|\hat h_k|}{1-\varepsilon}
\quad \Longleftrightarrow \quad
|\hat h_j|
\ge
\frac{1+\varepsilon}{1-\varepsilon}\,|\hat h_k|,
\quad \forall j>k.
\label{eq:conservative_order}
\end{equation}
In implementation, we sort users by $|\hat h_k|$ (ascending) and then verify \eqref{eq:conservative_order} for adjacent pairs; if a violation occurs, we group the ambiguous users into the same ``uncertainty cluster'' and keep a fixed internal order to avoid frequent order flips across realizations.
This conservative handling ensures that the intended gain ordering behind \eqref{eq:sic_reduced} is not invalidated by estimation errors.

\subsection{Conservative SINR Evaluation}
Moreover, SIC is not perfect in practical NOMA receivers due to channel estimation errors and non-ideal reconstruction. After subtracting the decoded signals, a small amount of uncanceled interference may remain. This residual SIC raises the effective interference floor and can noticeably degrade the minimum user SINR, which is particularly critical for max–min SINR designs. To account for imperfect CSI and residual SIC, we adopt a conservative SINR evaluation that guarantees SIC feasibility while avoiding overly optimistic assumptions of ideal SIC. Specifically,  we shrink the desired-signal power and inflate the interference terms by
\begin{equation}
\begin{aligned}
g_s(\varepsilon) &= (1-\varepsilon)^2,\\
g_i(\varepsilon) &= (1+\eta_i\varepsilon)^2,\\
g_r(\varepsilon) &= \eta_r\varepsilon,
\end{aligned}
\label{eq:robust_gains}
\end{equation}
where $\eta_i>0$ controls the interference inflation level and $\eta_r\ge 0$ quantifies the residual SIC leakage, and $g_s(\varepsilon)$, $g_i(\varepsilon)$, and $g_r(\varepsilon)$ respectively (i) downscale the desired-signal power to obtain a conservative lower bound, (ii) upscale the multiuser-interference power to form a conservative upper bound, and (iii) add a leakage term proportional to $\varepsilon$ to account for imperfect cancellation of previously decoded signals.
Define the conservative denominator
\begin{equation}
\begin{aligned}
D_k(\mathbf{x}_{\!p},\boldsymbol{\alpha})
&=
g_i(\varepsilon)\,P_T|h_k(\mathbf{x}_{\!p})|^2
\sum_{j=k+1}^{K}\alpha_j\\
&\quad +
g_r(\varepsilon)\,P_T|h_k(\mathbf{x}_{\!p})|^2
\sum_{j=1}^{k-1}\alpha_j
+\sigma^2,
\end{aligned}
\label{eq:sinr_den_conservative}
\end{equation}
and calculate
\begin{equation}
\mathrm{SINR}_k(\mathbf{x}_{\!p},\boldsymbol{\alpha}) =
\frac{g_s(\varepsilon)\,\alpha_k P_T |h_k(\mathbf{x}_{\!p})|^2}{D_k(\mathbf{x}_{\!p},\boldsymbol{\alpha})}.\label{eq:sinr_conservative}
\end{equation}
In the rest of the paper, SINR refers to the conservative SINR evaluation given by \eqref{eq:sinr_conservative} unless otherwise specified.

\subsection{Problem Formulation}
Based on the formulated conservative SINR, we aim to maximize the minimum user SINR subject to the NOMA power-allocation constraint and the PA geometry constraints, which yields the following problem:
\begin{align}
\max_{\mathbf{x}_{\!p},\,\boldsymbol{\alpha}}~~
& \Gamma_{\min}(\mathbf{x}_{\!p},\boldsymbol{\alpha})
\label{eq:opt_problem_final}\\
\text{s.t.}~~
& \alpha_k \ge 0,\quad \sum_{k=1}^{K}\alpha_k \le 1,
\label{eq:opt_constraint_power_noma_final}\\
& 0 \le x_n \le L,\quad n=1,\ldots,N,
\label{eq:opt_constraint_box_final}\\
& |x_n - x_{n-1}| \ge d_{\min},\quad n=2,\ldots,N,
\label{eq:opt_constraint_spacing_final}
\end{align}
where
\begin{equation}
\Gamma_{\min}(\mathbf{x}_{\!p},\boldsymbol{\alpha})
=
\min_{k} \mathrm{SINR}_k(\mathbf{x}_{\!p},\boldsymbol{\alpha}).
\label{eq:ob}
\end{equation}
The resulting problem is highly nonconvex due to the nonlinear dependence of $|h_k(\mathbf{x}_{\!p})|^2$ on the PA positions and the non-smooth minimum operator in $\Gamma_{\min}(\mathbf{x}_{\!p},\boldsymbol{\alpha})$.  Moreover, the SIC decoding order must remain feasible under CSI uncertainty, which further restricts the search space.  Therefore, we adopt PSO to handle the coupled design without requiring gradient information.

\section{Algorithm Design}
\label{sec:optimization}

This section presents a tailored PSO algorithm to solve problem \eqref{eq:opt_problem_final}.
The design variables are the PA positions $\mathbf{x}_{\!p}$ and the NOMA power allocation coefficient vector $\boldsymbol{\alpha}$.

\subsection{Fitness Evaluation}

Given a candidate $(\mathbf{x}_{\!p},\boldsymbol{\alpha})$, we:
(i) compute the PA--user channels $g_{k,n}$ via \eqref{eq:channel};
(ii) obtain the effective channels $h_k$ via \eqref{effective_channel} and set the nominal estimate $\hat h_k \triangleq h_k$, with CSI uncertainty captured only through $\varepsilon$ in \eqref{eq:conservative_order} and \eqref{eq:sinr_conservative};
(iii) construct a decoding order using the conservative evaluation \eqref{eq:conservative_order}, so that the ordering behind \eqref{eq:sic_reduced} is preserved under \eqref{eq:csi_error};
(iv) evaluate $\mathrm{SINR}_k$ using conservative SINR evaluation \eqref{eq:sinr_conservative};
(v) set the fitness as $\Gamma_{\min}(\mathbf{x}_{\!p},\boldsymbol{\alpha})$ defined in \eqref{eq:ob}.

To discourage candidate solutions that violate the conservative SIC ordering implied by \eqref{eq:conservative_order}, we add an order-violation penalty based on adjacent users in the decoding order used in the SINR evaluation. Define
\begin{equation}
[x]_+ \triangleq \max(x,0),
\label{eq:relu_def}
\end{equation}
and the adjacent-pair violation amount
\begin{equation}
v_k
\triangleq
\left[
\frac{1+\varepsilon}{1-\varepsilon}\,|\hat h_{k}|
-
|\hat h_{k+1}|
\right]_+,
\qquad k=1,\ldots,K-1,
\label{eq:order_violation}
\end{equation}
where the indices follow the decoding order adopted when evaluating \eqref{eq:sinr_conservative}.
We then use the penalized fitness
\begin{equation}
F(\mathbf{x}_{\!p},\boldsymbol{\alpha})
=
\Gamma_{\min}(\mathbf{x}_{\!p},\boldsymbol{\alpha})
-
\mu \sum_{k=1}^{K-1} v_k,
\label{eq:fitness_penalized}
\end{equation}
where $\mu>0$ is a penalty weight. In the PSO iterations, $F(\mathbf{x}_{\!p},\boldsymbol{\alpha})$ is used for comparing particles and updating the personal and global best states.

\subsection{PSO-Based Joint Optimization}

Each particle encodes $\boldsymbol{\theta}_i=[\mathbf{x}_{\!p,i}^T,\boldsymbol{\alpha}_i^T]^T$.
Feasibility requires \eqref{eq:opt_constraint_power_noma_final}--\eqref{eq:opt_constraint_spacing_final}.
Let $\boldsymbol{\theta}_i^{(t)}$ and $\boldsymbol{\nu}_i^{(t)}$ denote the position and velocity of particle $i$ at iteration $t$.
The standard PSO update is
\begin{equation}
\begin{aligned}
\boldsymbol{\nu}_i^{(t+1)}
&=
\omega\,\boldsymbol{\nu}_i^{(t)}
+c_1 r_{1,i}^{(t)}\big(\boldsymbol{\theta}_{i,\mathrm{best}}-\boldsymbol{\theta}_i^{(t)}\big)
+c_2 r_{2,i}^{(t)}\big(\boldsymbol{\theta}_{\mathrm{gbest}}-\boldsymbol{\theta}_i^{(t)}\big),\\
\widetilde{\boldsymbol{\theta}}_i^{(t+1)}
&=
\boldsymbol{\theta}_i^{(t)}+\boldsymbol{\nu}_i^{(t+1)},\\
\boldsymbol{\theta}_i^{(t+1)}
&=
\mathcal{P}\!\left(\widetilde{\boldsymbol{\theta}}_i^{(t+1)}\right),
\end{aligned}
\label{eq:pso_update_opt}
\end{equation}
where $\omega,c_1,c_2$ are PSO parameters and $r_{1,i}^{(t)},r_{2,i}^{(t)}\sim\mathcal{U}[0,1]$.
The projection step ensures feasibility after each update.
For geometry, a practical implementation is: (a) clip each $x_n$ to $[0,L]$ ;
(b) sort $\{x_n\}$ and enforce spacing by a forward pass $x_n\leftarrow \max(x_n, x_{n-1}+d_{\min})$ and a backward pass to keep $x_N\le L$;
(c) map back to the particle representation.
For power allocation, clip $\alpha_k\ge 0$ and project onto the simplex $\sum_k\alpha_k\le 1$.
For each particle, computing $\{g_{k,n}\}$ costs $\mathcal{O}(KN)$ and evaluating $\{\mathrm{SINR}_k\}$ costs $\mathcal{O}(K)$ once $\{|h_k|^2\}$ are available.
Thus, each iteration costs $\mathcal{O}(KN)$ per particle and the overall complexity is
\begin{equation}
\mathcal{O}\!\left(N_{\mathrm{part}}\,T_{\max}\,K\,N\right).
\end{equation}
The details of the proposed PSO algorithm are summarized in Algorithm \ref{alg:robust_pso}.
\begin{algorithm}[t]
\caption{Robust PSO for joint Optimization of PA Positions and NOMA Power Allocation Coefficients}
\label{alg:robust_pso}
\begin{algorithmic}[1]
\Require $K,N,L,P_T,d_{\min},\sigma^2,\lambda,\beta,\alpha,\varepsilon,\eta_i,\eta_r,\mu,$
\Statex \hspace{\algorithmicindent} $N_{\mathrm{part}},T_{\max},\omega,c_1,c_2$
\Ensure $(\mathbf{x}_{\!p}^{\star},\boldsymbol{\alpha}^{\star})$
\State Initialize particles $\{\boldsymbol{\theta}_i^{(0)},\boldsymbol{\nu}_i^{(0)}\}_{i=1}^{N_{\mathrm{part}}}$;
$\boldsymbol{\theta}_i^{(0)}\!\leftarrow\!\mathcal{P}(\boldsymbol{\theta}_i^{(0)})$
\For{$i=1$ to $N_{\mathrm{part}}$}
    \State Compute $F_i=\Gamma_{\min}(\boldsymbol{\theta}_i^{(0)})$ using
    \eqref{eq:channel}, \eqref{effective_channel},
    \eqref{eq:conservative_order}, \eqref{eq:sinr_den_conservative}, \eqref{eq:sinr_conservative}, \eqref{eq:ob}
    \State Set $\hat h_k \leftarrow h_k$; sort by $|\hat h_k|$ and compute $V_i \triangleq \sum_{k=1}^{K-1} v_k$ via \eqref{eq:order_violation}
    \State Penalize fitness: $F_i \leftarrow F_i - \mu V_i$
    \State $\boldsymbol{\theta}_{i,\mathrm{best}}\!\leftarrow\!\boldsymbol{\theta}_i^{(0)}$,\quad $F_{i,\mathrm{best}}\!\leftarrow\!F_i$
\EndFor
\State $\boldsymbol{\theta}_{\mathrm{gbest}}\!\leftarrow\!\arg\max_i F_{i,\mathrm{best}}$
\For{$t=0$ to $T_{\max}-1$}
    \For{$i=1$ to $N_{\mathrm{part}}$}
        \State Update $\boldsymbol{\nu}_i^{(t+1)}$, $\widetilde{\boldsymbol{\theta}}_i^{(t+1)}$ via \eqref{eq:pso_update_opt}
        \State $\boldsymbol{\theta}_i^{(t+1)}\!\leftarrow\!\mathcal{P}\!\left(\widetilde{\boldsymbol{\theta}}_i^{(t+1)}\right)$
        \State Compute $F_i=\Gamma_{\min}(\boldsymbol{\theta}_i^{(t+1)})$ using
        \eqref{eq:channel}, \eqref{effective_channel},
        \eqref{eq:conservative_order}, \eqref{eq:sinr_den_conservative}, \eqref{eq:sinr_conservative}, \eqref{eq:ob}
        \State Set $\hat h_k \leftarrow h_k$; sort by $|\hat h_k|$ and compute $V_i \triangleq \sum_{k=1}^{K-1} v_k$ via \eqref{eq:order_violation}
        \State Penalize fitness: $F_i \leftarrow F_i - \mu V_i$
        \If{$F_i>F_{i,\mathrm{best}}$}
            \State $\boldsymbol{\theta}_{i,\mathrm{best}}\!\leftarrow\!\boldsymbol{\theta}_i^{(t+1)}$,\quad $F_{i,\mathrm{best}}\!\leftarrow\!F_i$
        \EndIf
    \EndFor
    \State $\boldsymbol{\theta}_{\mathrm{gbest}}\!\leftarrow\!\arg\max_i F_{i,\mathrm{best}}$
\EndFor
\State \Return $(\mathbf{x}_{\!p}^{\star},\boldsymbol{\alpha}^{\star})$ from $\boldsymbol{\theta}_{\mathrm{gbest}}$
\end{algorithmic}
\end{algorithm}
%
\section{Numerical Results}
\label{sec:results}

This section presents numerical results for the downlink NOMA-enabled PASS. All results are averaged over 1000 independent channel and blockage realizations.


Four schemes are compared: \textbf{Robust-PSO} (robust PSO optimizing PA positions and NOMA power allocation coefficients under imperfect CSI), \textbf{Non-Robust-PSO} (PSO with perfect CSI), \textbf{Random} (random feasible PA positions and power allocation), and \textbf{Uniform} (fixed PA positions with uniform power allocation). Table~\ref{tab:default_params} lists the default configuration used unless otherwise stated.

\begin{table}[t]
\centering
\caption{Default Simulation Configuration}
\label{tab:default_params}
\begin{tabular}{l l l}
\hline
\textbf{Category} & \textbf{Parameter (Symbol)} & \textbf{Default Value} \\
\hline
System scale
& Number of users ($K$) & $3$ \\
& Number of PAs ($N$) & $5$ \\
& Waveguide length ($L$) & $10$ m \\
& Service area size  & $10\,\mathrm{m} \times 10\,\mathrm{m}$ \\
\hline
Channel
& Carrier frequency ($f_c$) & $28$ GHz \\
\hline
Robust design
& CSI error ratio ($\varepsilon$) & $0.1$ \\
& Interference inflation ($\eta_i$) & $0.5$ \\
& Residual SIC factor ($\eta_r$) & $0.2$ \\
\hline
Fitness penalty
& Order-violation weight ($\mu$) & $1.0$ \\
\hline
Blockage model
& Minimum transmission coefficient ($\beta$) & $0.1$ \\
& Attenuation rate ($\alpha$) & $2.0$ \\
\hline
\end{tabular}
\end{table}


\begin{figure}[t]
	\vspace{-0.3cm}
\centering
\includegraphics[width=0.9\columnwidth]{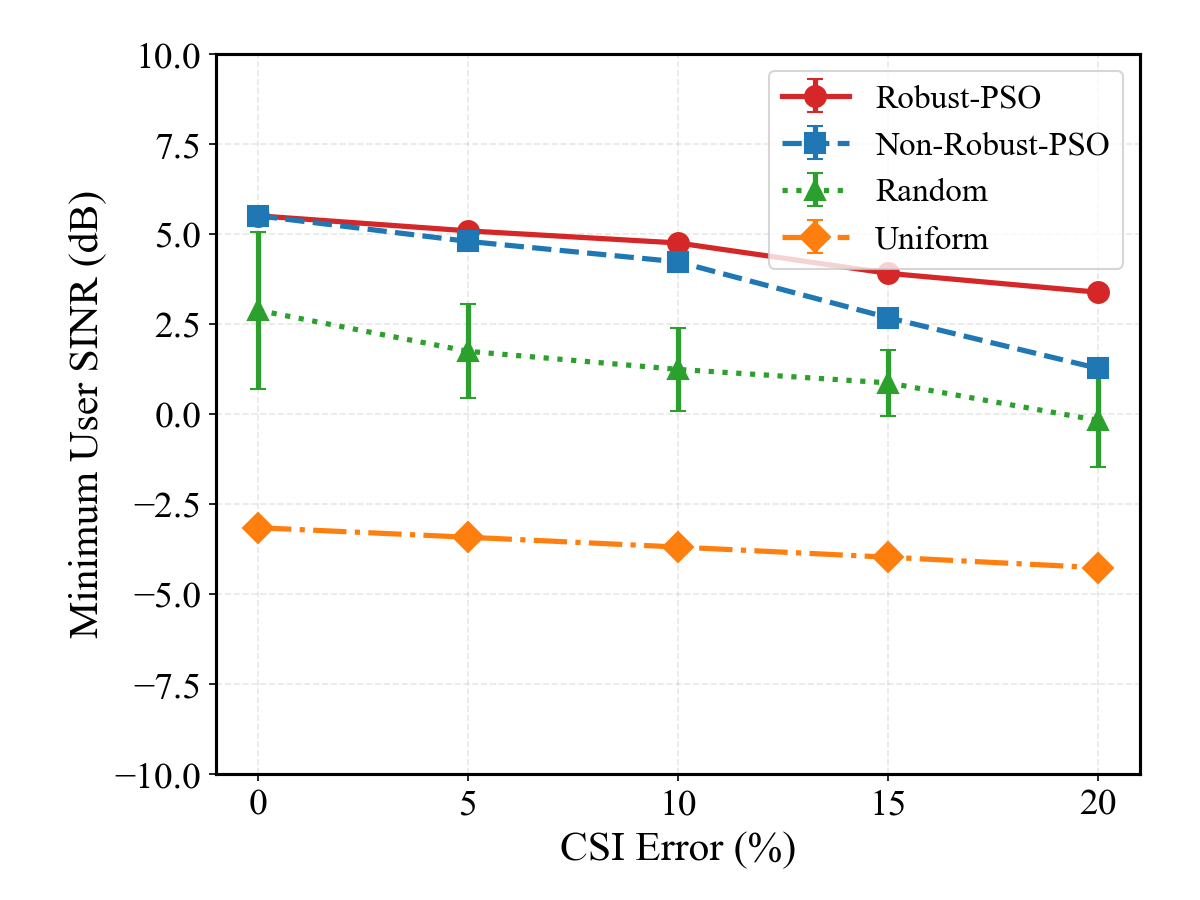}
\vspace{-0.4cm}
\caption{Minimum user SINR versus CSI error ratio $\varepsilon$ for different schemes.}
\label{fig:csi_error}
\end{figure}

Fig.~\ref{fig:csi_error} shows that Robust-PSO maintains a higher minimum user SINR than Non-Robust-PSO as $\varepsilon$ increases, while Random and Uniform remain much worse. The Robust-to-Uniform improvement remains large across all CSI error levels and averages $7.70$ dB over $\varepsilon\in[0,20\%]$. The Robust-to-Non-Robust gap is small when $\varepsilon=0$ and widens as $\varepsilon$ increases, which confirms that optimizing with perfect CSI alone can lead to solutions that are sensitive to estimation mismatch. In contrast, the robust fitness guides the search toward configurations that preserve fairness when the realized channel gains deviate within the uncertainty set.



\begin{figure}[t]
		\vspace{-0.4cm}
\centering
\includegraphics[width=0.9\columnwidth]{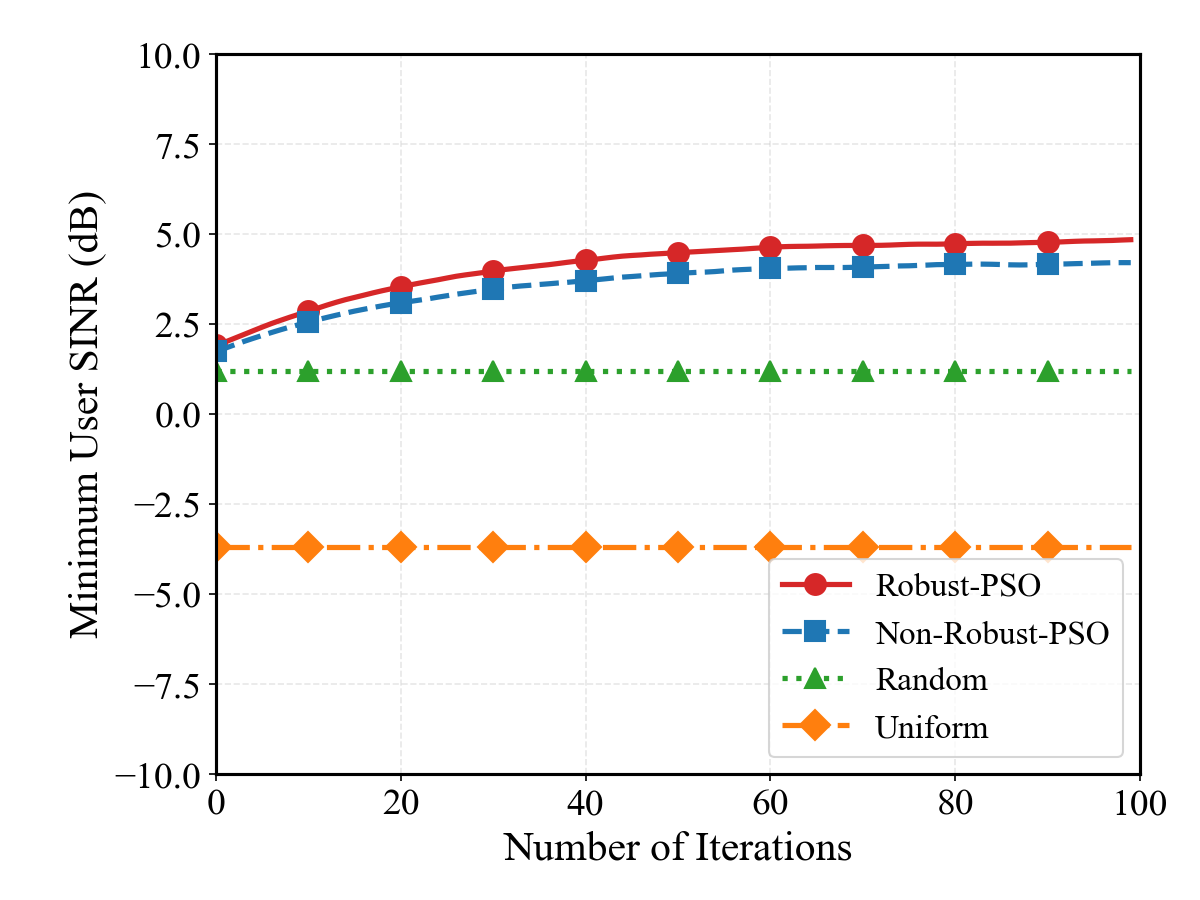}
\vspace{-0.4cm}
\caption{Minimum user SINR versus PSO iteration index for different schemes.}
\label{fig:conv_all}
\end{figure}

\begin{figure}[!t]
		\vspace{-0.3cm}
\centering
\includegraphics[width=0.9\columnwidth]{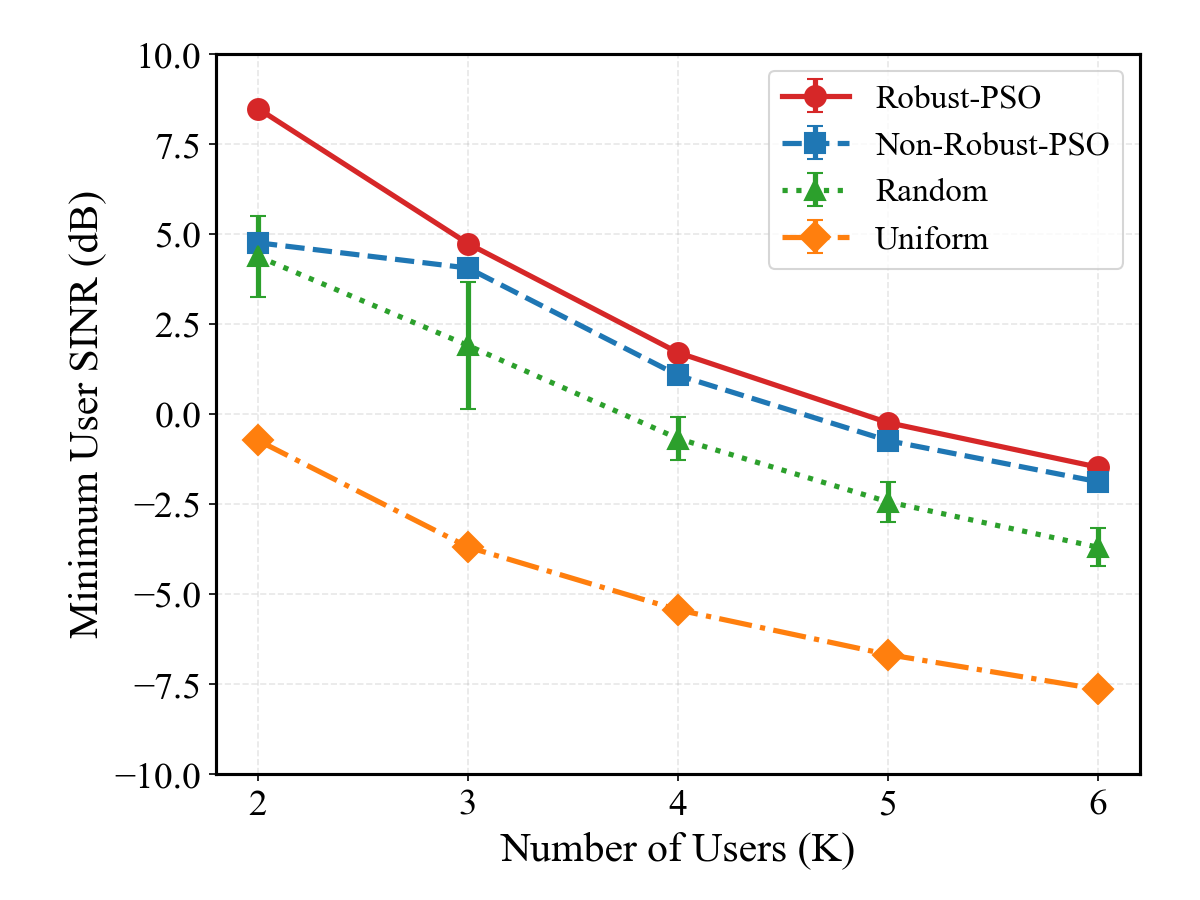}
\vspace{-0.4cm}
\caption{Minimum user SINR versus the number of users $K$.}
\label{fig:users_min_sinr}
\end{figure}

In Fig.~\ref{fig:conv_all}, both PSO-based designs improve the minimum user SINR rapidly in early iterations and then approach a steady value, which indicates stable convergence under the selected swarm size and iteration budget. 
Robust-PSO converges to a higher fairness level than Non-Robust-PSO because its robustness-aware fitness explicitly accounts for imperfect CSI and residual SIC, thereby discouraging solutions that are only optimal under optimistic channel estimates. 
Non-Robust-PSO uses a nominal-CSI fitness, so it can favor configurations that appear strong under estimated channels but lose fairness when the realized gains deviate, which explains the consistent gap at the plateau.

In Fig.~\ref{fig:users_min_sinr}, the minimum user SINR decreases as $K$ increases for all schemes due to stronger multiuser interference and stricter power sharing under NOMA. 
Robust-PSO remains best across the tested range because PA positioning reshapes near-field channel disparities and power allocation further protects the weakest user, which jointly improves the effective gain ordering and the worst-user SINR. 
Random and Uniform cannot exploit these spatial and power-domain degrees of freedom, so their minimum user SINR degrades faster as the system load increases.


\section{Conclusion}
This paper investigated a NOMA-enabled PASS under blockage and imperfect CSI. A channel model incorporating soft blockage and imperfect CSI was developed, based on which a max–min SINR optimization problem was formulated to jointly optimize PA positions and NOMA power allocation coefficients. To ensure the feasibility of SIC under CSI errors, a conservative SINR evaluation was introduced, and the resulting non-convex problem was efficiently solved using a tailored PSO algorithm. Simulation results demonstrated that the proposed design significantly improves the minimum user SINR compared with fixed-antenna systems and other baselines under moderate blockage and imperfect CSI. 
\bibliographystyle{IEEEtran}
\bibliography{IEEEabrv,MyBibliography}

\begin{thebibliography}{10}
\providecommand{\url}[1]{#1}
\csname url@samestyle\endcsname
\providecommand{\newblock}{\relax}
\providecommand{\bibinfo}[2]{#2}
\providecommand{\BIBentrySTDinterwordspacing}{\spaceskip=0pt\relax}
\providecommand{\BIBentryALTinterwordstretchfactor}{4}
\providecommand{\BIBentryALTinterwordspacing}{\spaceskip=\fontdimen2\font plus
\BIBentryALTinterwordstretchfactor\fontdimen3\font minus
  \fontdimen4\font\relax}
\providecommand{\BIBforeignlanguage}[2]{{%
\expandafter\ifx\csname l@#1\endcsname\relax
\typeout{** WARNING: IEEEtran.bst: No hyphenation pattern has been}%
\typeout{** loaded for the language `#1'. Using the pattern for}%
\typeout{** the default language instead.}%
\else
\language=\csname l@#1\endcsname
\fi
#2}}
\providecommand{\BIBdecl}{\relax}
\BIBdecl

\bibitem{suzuki2022pinching}
H.~O.~Y. Suzuki and K.~Kawai, ``{Pinching Antenna Using a Dielectric Waveguide
  as an Antenna},'' \emph{Technical J.}, vol.~23, no.~3, pp. 5--12, 2022.

\bibitem{11036558}
Z.~Ding and H.~Vincent~Poor, ``{LoS Blockage in Pinching-Antenna Systems: Curse
  or Blessing?}'' \emph{IEEE Wireless Communications Letters}, vol.~14, no.~9,
  pp. 2798--2802, 2025.

\bibitem{10945421}
Z.~Ding, R.~Schober, and H.~Vincent~Poor, ``{Flexible-Antenna Systems: A
  Pinching-Antenna Perspective},'' \emph{IEEE Transactions on Communications},
  vol.~73, no.~10, pp. 9236--9253, 2025.

\bibitem{10912473}
K.~Wang, Z.~Ding, and R.~Schober, ``{Antenna Activation for NOMA Assisted
  Pinching-Antenna Systems},'' \emph{IEEE Wireless Communications Letters},
  vol.~14, no.~5, pp. 1526--1530, 2025.

\bibitem{11016750}
X.~Xie, F.~Fang, Z.~Ding, and X.~Wang, ``{A Low-Complexity Placement Design of
  Pinching-Antenna Systems},'' \emph{IEEE Communications Letters}, vol.~29,
  no.~8, pp. 1784--1788, 2025.

\bibitem{11029492}
Z.~Zhou, Z.~Yang, G.~Chen, and Z.~Ding, ``{Sum-Rate Maximization for
  NOMA-Assisted Pinching-Antenna Systems},'' \emph{IEEE Wireless Communications
  Letters}, vol.~14, no.~9, pp. 2728--2732, 2025.

\bibitem{xu2025qosawarenomadesigndownlink}
\BIBentryALTinterwordspacing
Y.~Xu, Z.~Ding, D.~Cai, and V.~W.~S. Wong, ``{QoS-Aware NOMA Design for
  Downlink Pinching-Antenna Systems},'' 2025. [Online]. Available:
  \url{https://arxiv.org/abs/2504.13723}
\BIBentrySTDinterwordspacing

\bibitem{11186151}
S.~Mohammadzadeh, K.~Cumanan, C.~Li, and Z.~Ding, ``{Efficient Downlink Power
  Allocation for NOMA-Based Pinching-Antenna Systems},'' \emph{IEEE Wireless
  Communications Letters}, vol.~14, no.~12, pp. 4187--4191, 2025.

\bibitem{gan2025jointbeamformingnomaassisted}
\BIBentryALTinterwordspacing
D.~Gan, X.~Xu, J.~Zuo, X.~Ge, and Y.~Liu, ``{Joint Beamforming for NOMA
  Assisted Pinching Antenna Systems (PASS)},'' 2025. [Online]. Available:
  \url{https://arxiv.org/abs/2506.03063}
\BIBentrySTDinterwordspacing

\bibitem{xue2025resourceallocationpinchingantennasystems}
\BIBentryALTinterwordspacing
S.~Xue, J.~Zhao, K.~Cai, X.~Mu, Z.~Xiao, and Y.~Liu, ``{Resource Allocation for
  Pinching-Antenna Systems (PASS)-enabled NOMA Communications},'' 2025.
  [Online]. Available: \url{https://arxiv.org/abs/2512.03502}
\BIBentrySTDinterwordspacing

\end{thebibliography}
\end{document}